\documentclass[aps,pr,superscriptaddress,preprintnumbers,nofootinbib,10pt]{revtex4-2}

\usepackage{multirow}
\usepackage{amsmath}
\usepackage{amssymb}
\usepackage[dvipdf,dvips]{graphicx}
\usepackage{color}
\usepackage{hyperref}
\usepackage{url}
\usepackage{slashed}
\usepackage{subfigure}
\usepackage[usenames,dvipsnames]{xcolor}
\usepackage{amsmath}
\usepackage{amsfonts}
\usepackage{float} 
\usepackage{amssymb}
\usepackage{epsfig}
\usepackage{graphics}
\usepackage{euscript}
\usepackage{slashed}
\usepackage{epstopdf}
\usepackage[utf8]{inputenc}
\allowdisplaybreaks
\usepackage[normalem]{ulem}
\usepackage{pifont}
\usepackage{dsfont}
\usepackage{MnSymbol}
\usepackage{verbatim}
\usepackage{graphicx}
\usepackage{latexsym}
\usepackage{tikz-feynman}
\usepackage{bbold}

\begin{document}

	\title{{\bf \Large  On a class of bounded Hermitian operators for the Bell-CHSH inequality in Quantum Field Theory }}
	
	\vspace{1cm}
	
	\author{M. S.  Guimaraes}\email{msguimaraes@uerj.br} \affiliation{UERJ $–$ Universidade do Estado do Rio de Janeiro,	Instituto de Física $–$ Departamento de Física Teórica $–$ Rua São Francisco Xavier 524, 20550-013, Maracanã, Rio de Janeiro, Brazil}
	
	\author{I. Roditi} \email{roditi@cbpf.br} \affiliation{CBPF $-$ Centro Brasileiro de Pesquisas Físicas, Rua Dr. Xavier Sigaud 150, 22290-180, Rio de Janeiro, Brazil } \affiliation{Institute for Theoretical Physics, ETH Zürich, 8093 Zürich, Switzerland} 
	
	\author{S. P. Sorella} \email{silvio.sorella@fis.uerj.br} \affiliation{UERJ $–$ Universidade do Estado do Rio de Janeiro,	Instituto de Física $–$ Departamento de Física Teórica $–$ Rua São Francisco Xavier 524, 20550-013, Maracanã, Rio de Janeiro, Brazil}

	\begin{abstract}
		
	The violation of the Bell-CHSH inequality in a relativistic scalar Quantum Field Theory is analysed by means of a set of bounded Hermitian operators constructed out of the unitary Weyl operators. These operators allow for both analytic and numerical approaches. While the former relies on the modular theory of Tomita-Takesaki, the latter is devised through an explicit construction of the test functions needed for the localization of the aforementioned operators. The case of causal tangent diamonds in $1+1$ Minkowski spacetime is scrutinized. 	
		
		\end{abstract}

	\maketitle

	\section{Introduction}\label{intro}
	
	Since the pioneering work of Summers-Werner \cite{Summers:1987fn,Summ,Summers:1987ze}, the study of the Bell-CHSH inequality  \cite{Bell:1964kc,Clauser:1969ny} is a seminal tool for unravelling many aspects of the unavoidable presence of the entanglement in relativistic Quantum Field Theory  \cite{Witten:2018zxz, Guimaraes:2024mmp}. \\\\As pointed out in \cite{Summers:1987fn,Summ,Summers:1987ze,Llandau}, in a Quantum Field Theory, the vacuum state itself $|0\rangle$ is highly entangled. This feature has been established by a series of Theorems \cite{Summers:1987fn,Summ,Summers:1987ze} stating that, in the case of causal complementary wedge regions as well as for causal tangent diamonds in Minkowski spacetime, the Bell-CHSH inequality in the vacuum state can be maximally violated, attaining Tsirelson's bound \cite{tsi1}, namely: $2\sqrt{2}$. It is worth underlining that these remarkable results hold for free quantum fields, both of Bose and Fermi type. \\\\The proof of the Theorems makes use of fundamental tools of Quantum Field Theory, such as: the Haag-Kastler algebraic formulation \cite{Haag:1992hx}, the Reeh-Schlieder \cite{Reeh:1961ujh} and the Bisognano-Wichmann \cite{BW} Theorems, the von Neumann algebras \cite{Sorce:2023fdx}, the modular theory of Tomita-Takesaki \cite{TT,Guido:2008jk}. \\\\On the other hand, the translation of these Theorems in terms of field models allowing for an explicit computational framework of the violation of the Bell-CHSH inequality in the vacuum state is still an open issue, see, for instance, \cite{Dudal:2023mij,DeFabritiis:2023tkh,DeFabritiis:2024jfy,Guimaraes:2024lqf,Guimaraes:2024xtj,Guimaraes:2025xij}. \\\\Basically, one is looking at a correlation function of the following type 
\begin{equation} 
\langle 0| \; {\cal C} \;|0\rangle = \langle 0|\; (A(f) + A(f'))B(g) + (A(f) -A(f'))B(g') \;|0\rangle \;, \label{corrf}
\end{equation}
where $(A(f),A(f'))$, $(B(g),B(g'))$ are bounded Hermitian field operators whose norm is less or equal to $1$, {\it i.e}
\begin{equation} 
||A(f)|| \le 1 \;, \qquad ||A(f')|| \le 1\;, \qquad ||B(g)|| \le 1 \;, \qquad ||B(g')||\le 1  \;. \label{le1}
\end{equation} 
The quantities $(f,f')$ and $(g,g')$ are smooth test functions with compact support, needed to properly localize the operators $(A,B)$ in the desired regions of the Minkowski spacetime. The supports of $(f,f')$ and $(g,g')$ are space-like 
\begin{equation} 
supp(f,f') \;\;\;{\rm space-like} \;\;\; supp(g,g')  \nonumber
\end{equation} 
As a consequence, from causality \cite{Haag:1992hx}, it follows that 
\begin{equation} 
[A,B]=0 \;. \label{caus}
\end{equation}
It is customary to refer to $(A,A')$ as to Alice's operators and to $(B,B')$ as Bob's operators. \\\\To be concrete, in what follows we shall consider a free massive scalar field  $\varphi(t,x)$ in $1+1$ Minkowski spacetime	 
\begin{equation} \label{qf}
		\varphi(t,x) = \int \! \frac{d k}{2 \pi} \frac{1}{2 \omega_k} \left( e^{-ik_\mu x^\mu} a_k + e^{ik_\mu x^\mu} a^{\dagger}_k \right) \;, 
	\end{equation} 
where $\omega_k  = k^0 = \sqrt{k^2 + m^2}$ and 
	\begin{align}
		[a_k, a^{\dagger}_q] &= (2\pi) \, 2\omega_k \, \delta(k - q), \\ \nonumber 
		[a_k, a_q] &= [a^{\dagger}_k, a^{\dagger}_q] = 0. 
	\end{align}
Due to its singular nature, $\varphi(t,x)$ has to be treated as an operator valued distribution  \cite{Haag:1992hx} and, as such, it needs to be smeared out, see Appendix \eqref{appA}:
\begin{equation} 
\varphi(h) = \int d^2x\; \varphi(t,x) h(t,x) \;, \label{smm}
\end{equation}
where $h(t,x)$ stands for a generic smooth test function with compact support. Equation \eqref{smm} can be rewritten as 
\begin{equation} 
\varphi(h) = a_h + a^{\dagger}_h \;, \qquad a_h = \int \! \frac{d k}{2 \pi} \frac{1}{2 \omega_k} h^{*}(\omega_k ,k) a_k \;, \qquad a_h^{\dagger} = \int \! \frac{d k}{2 \pi} \frac{1}{2 \omega_k} h(\omega_k ,k) a_k^{\dagger} \;, \label{phih}
\end{equation}
and 
\begin{equation} 
h(\omega_k,k) = \int d^2x\; e^{i(\omega_k t - kx)} \; h(t,x) \;. \label{emb}
\end{equation}
Thus, the short handed notation $(A(f),A(f'))$, $(B(g),B(g')$ means: $(A(\varphi(f)), A(\varphi(f')))$, $(B(\varphi(g)), B(\varphi(g')))$. \\\\The Bell-CHSH inequality is said to be violated in the vacuum state $|0\rangle$ whenever 
\begin{equation} 
2 < \Big| \langle 0|\; {\cal C} \;|0\rangle \Big| \le 2 \sqrt{2}\;.  \label{vbc}
\end{equation}
In the last years, we have undertaken much efforts \cite{Dudal:2023mij,DeFabritiis:2023tkh,DeFabritiis:2024jfy,Guimaraes:2024lqf,Guimaraes:2024xtj,Guimaraes:2025xij} towards the achievement of a computational setup for expression \eqref{vbc}. This endeavour has required the facing of the explicit choice of the test functions $(f,f')$, $(g,g')$ as well as the formulation of a numerical framework for the evaluation of the complex spacetime integrals needed for a Bell test in Quantum Field Theory. \\\\In the present work we aim at presenting new results for the Bell-CHSH correlation function of eq.\eqref{corrf}, as enlisted below, namely: 
\begin{itemize} 
\item a class of Hermitian bounded operators which can be evaluated by making use of the unitary Weyl operators 
  \cite{DeFabritiis:2023tkh} 
\begin{align}
		W_f = e^{i \varphi(f)}= e^{i(a_f + a^\dagger_f)} \;, \qquad W_f^{\dagger} W_f = W_f W_f^\dagger = 1 \;. \qquad W^{\dagger}_f = W_{-f} \;, \label{ww}
	\end{align}
is constructed. These operators turn out to be treated both analytically, via the Tomita-Takesaki modular theory \cite{TT,Guido:2008jk}, as well as numerically. 
\item a quite helpful parametrization of the Bell-CHSH correlation function is introduced, allowing for an efficient comparison between the Tomita-Takesaki theoretical approach and the numerical setup, leading to rather large violations of the Bell-CHSH inequality.
\end{itemize}
The work is organized as follows. In Sect.\eqref{HOp} we outline the construction of the aforementioned operators. In Sect.\eqref{TTmd} we address the issue of the choice of the test functions, both from the Tomita-Takesaki and the numerical side. In Sect.\eqref{nm} we present the violations of the Bell-CHSH inequality obtained by considering three different examples of bounded Hermitian operators. Sect.\eqref{Ccc} collects our conclusion. Finally, in Appendix \eqref{appA}, we give a short summary of the canonical quantization of the massive scalar field in $1+1$ Minkowski spacetime.

\section{Construction of a class of bounded Hermitian operators}\label{HOp}	

In order to construct a suitable class of Hermitian operators, we shall consider real bounded functions $\{ \sigma(x) \}$ with known Fourier transformation: 
\begin{equation} 
\sigma(x) = \int dk\; e^{ikx}\; {\hat \sigma}(k) \;, \label{ftk}
\end{equation}
where ${\hat \sigma}(k)$ is known in closed form. Examples of $\{ \sigma(x)\}$ are given by 
\begin{eqnarray} 
\frac{1}{cosh(x)} & \Rightarrow &  {\hat \sigma}(k) = \frac{1}{2} sech\left( \frac{\pi k}{2} \right) \;, \nonumber \\
\frac{1}{1+x^2} & \Rightarrow &  {\hat \sigma}(k) = \frac{1}{2} e^{-|k|} \;, \nonumber \\
e^{-x^2} & \Rightarrow & {\hat \sigma}(k) = \frac{1}{\sqrt{\pi}} e^{-k^2} \;. \label{ftkex}
\end{eqnarray}
Therefore, from the functional calculus \cite{BR}  we write 
\begin{eqnarray} 
A_1(f) & =&  \frac{1}{cosh(\varphi(f))}  =  \frac{1}{2} \int_{-\infty}^{\infty} dk\; sech\left( \frac{\pi k}{2} \right)  \; e^{ik \varphi(f)} \;, \nonumber \\
A_2(f) & = & \frac{1}{1+ \varphi^2(f)}  =  \frac{1}{2} \int_{-\infty}^{\infty} dk\; e^{-|k|} \; e^{ik \varphi(f)} \;, \nonumber \\
A_3(f) & = & e^{- \varphi^2(f)} = \frac{1}{\sqrt{\pi}} \int_{-\infty}^{\infty} dk\; e^{-k^2} \; e^{ik \varphi(f)} \;. \label{opftA}
\end{eqnarray}
Similarly, for $(B_1(g), B_2(g), B_3(g))$, one has
\begin{eqnarray} 
B_1(g) & =&  \frac{1}{cosh(\varphi(g))}  =  \frac{1}{2} \int_{-\infty}^{\infty} dk\; sech\left( \frac{\pi k}{2} \right)  \; e^{ik \varphi(g)} \;, \nonumber \\
B_2(g) & = & \frac{1}{1+ \varphi^2(g)}  =  \frac{1}{2} \int_{-\infty}^{\infty} dk\; e^{-|k|} \; e^{ik \varphi(g)} \;, \nonumber \\
B_3(g) & = & e^{- \varphi^2(g)} = \frac{1}{\sqrt{\pi}} \int_{-\infty}^{\infty} dk\; e^{-k^2} \; e^{ik \varphi(g)} \;. \label{opftB}
\end{eqnarray}
All these operators are manifestly bounded and Hermitian. They will be  object of a detailed analysis. \\\\The possibility of evaluating the correlation functions of the operators in eqs.\eqref{opftA},\eqref{opftB} stems from the fact that they are expressed as continuous combinations of the Weyl operators. For instance, for the correlation function $\langle 0|\; A_1(f) B_1(g) \;|0\rangle$, we have 
\begin{equation} 
\langle 0|\; A_1(f) B_1(g) \;|0\rangle = \frac{1}{4} \int_{-\infty}^{\infty} dk dp \;sech\left( \frac{\pi k}{2} \right) \;sech\left( \frac{\pi k}{2} \right) \;
\langle 0|\; e^{ik \varphi(f)}\; e^{ip\varphi(g)} \;|0\rangle \;. \label{ccww}
\end{equation} 
Moreover, employing the Baker-Campbell-Hausdorff formula, for the correlation function of the Weyl operators, we have 
\begin{equation} 
\langle 0|\; e^{ik\varphi(f)} \; e^{ip \varphi(g)} \;|0\rangle = e^{-\frac{1}{2} ||kf +pg||^2} \;, \label{wwcc}
\end{equation}
where $|| \;\cdot \;||^2$ is the norm induced by the Lorentz invariant inner product $\langle f|g\rangle$, see App.\eqref{appA}:
\begin{equation} 
\langle f |g \rangle =  \int \! \frac{d k}{2 \pi} \frac{1}{2 \omega_k} f^{*}(\omega_k,k) g(\omega_k,k) = \frac{i}{2} \Delta_{PJ}(f,g) + H(f,g) \;, \label{iinn}
\end{equation}
with $\Delta_{PJ}(f,g)$ and $H(f,g)$ being, respectively, the Pauli-Jordan and the Hadamard smeared distributions, App.\eqref{appA}. \\\\Taking into account that the supports of $f$ and $g$ are spacelike, it follows that 
\begin{equation} 
\langle 0|\; e^{ik\varphi(f)} \; e^{ip \varphi(g)} \;|0\rangle = e^{-\frac{1}{2} (k^2 H(f,f) + p^2 H(g,g) + 2kp H(f,g) )} \;. \label{HHH}
\end{equation}
Finally 
\begin{equation} 
\langle 0|\; A_1(f) B_1(g) \;|0\rangle = \frac{1}{4} \int_{-\infty}^{\infty} dk dp \;sech\left( \frac{\pi k}{2} \right) \;sech\left( \frac{\pi k}{2} \right) \;e^{-\frac{1}{2} (k^2 H(f,f) + p^2 H(g,g) + 2kp H(f,g) )} \;. \label{HHH1}
\end{equation}
Similar expressions hold for the operators $(A_2,A_3)$ and corresponding $(B_2,B_3)$. Equation \eqref{HHH1} will be the main tool of our investigation. As we shall see, it allows for a very natural and helpful comparison between the analytic approach provided by the Tomita-Takesaki modular theory and the numerical setup.

\section{The Bell-CHSH inequality and the Tomita-Takesaki modular theory}\label{TTmd}

Let us begin by providing a short summary of the Tomita-Takesaki modular theory. Let ${\cal M}(O)$ be the set of test functions with compact support contained in the open region of the Minkowski spacetime $O$: 
\begin{equation} 
{\cal M}(O) = \{ h, \; supp(h) \subset O \} \;. \label{MMOO}
\end{equation} 
For further use, it is helpful to introduce the symplectic complement ${\cal M}'(O)$
\begin{equation} 
{\cal M}'(O) = \{ g, \; \Im\langle f|g\rangle = \frac{1}{2} \Delta_{PJ}(f,g) =0, \forall f \in {\cal M}(O) \} \;. \label{MMss}
\end{equation}  
In other words, ${\cal M}'(O)$ is the set of test functions whose supports are spacelike with respect to ${\cal M}(O)$. \\\\One introduces the von-Neumann algebra ${\cal A}({\cal M})$ obtained by means of the Weyl operators 
\begin{equation} 
{\cal A(}{\cal M}) = \{ h\in {\cal M},\; W_h= e^{i\varphi(h)}, \; supp(h) \subset O \}^{"} \;, \label{vnam}
\end{equation}
where the symbol ${"}$ means the bicommutant \cite{Guimaraes:2024mmp}. We remind here that the commutant ${\cal A}'({\cal M})$ is defined as the set of all elements which commute with each element of ${\cal A}({\cal M})$, {\it i.e.}
\begin{equation} 
{\cal A}'({\cal M}) = \{ w', w'w = w w'\;, \forall w \in {\cal A}({\cal M}) \} \;. \label{cctt}
\end{equation}
Therefore, for the bicommutant, one has 
\begin{equation}
{\cal A}^{"}({\cal M}) = ( {\cal A}'({\cal M}))^{'} \;. \label{bbcc}
\end{equation}
Also, the Weyl operators $W_h$ obey the following relation
\begin{eqnarray} 
W_h \; W_{h'} & = & e^{-\frac{i}{2} \Delta_{PJ}(h,h')} W_{h+h'} \;, \nonumber \\
\end{eqnarray}
An important tool in order to introduce the Tomita-Takesaki modular theory is the Reeh-Schlieder Theorem \cite{Haag:1992hx}, stating that the vacuum state $|0\rangle$ is cyclic and separating for the von Neumann algebra \eqref{vnam}. The Tomita-Takesaki construction relies on the introduction of an anti-linear unbounded operator $S$ \cite{TT,Guido:2008jk} acting on the von Neumann algebra ${\cal A}({\cal M})$ as 
\begin{equation} 
S \; w\;|0\rangle = w^{\dagger}\; |0\rangle \;, \qquad \forall w \in {\cal A}({\cal M}) \;. \label{STT}
\end{equation}
From the above definition, it follows that 
\begin{equation} 
S\; |0\rangle =|0\rangle \;, \qquad S^2=1 \;. \label{STTp}
\end{equation}
Further, one introduces the polar decomposition 
\begin{equation} 
S = J\; \Delta^{1/2} \;. \label{pold}
\end{equation} 
The operator $\Delta$, called the modular operator, is positive and self-adjoint, while $J$, called the modular conjugation, is anti-unitary. The operators $(J,\Delta)$ turn out to obey the following relations: 
\begin{eqnarray} 
J^2 & = & 1 \;, \qquad J^\dagger=J \;, \nonumber \\
J \Delta^{1/2} J & =& \Delta^{-1/2} \;, \qquad S^\dagger = J \Delta^{-1/2} \;, \qquad \Delta = S^\dagger S \;. \label{SDJ}
\end{eqnarray}
The Tomita-Takesaki Theorem \cite{TT} states that 
\begin{itemize} 
\item 
\begin{equation} 
J {\cal A}({\cal M}) J = {\cal A}'({\cal M}) \;,  \label{TTT1}
\end{equation} 
\item there exists a one-parameter family of unitary operators $\Delta^{it}$, $t \in \mathbb{R}$, such that 
\begin{equation} 
\Delta^{it} \; {\cal A}({\cal M}) \; \Delta^{-it} = {\cal A}({\cal M}) \;, \qquad \Delta^{it} = e^{it \log(\Delta)} \;, \label{TTT2}. 
\end{equation} 
\end{itemize}
The first equation tells us that the modular conjugation $J$ maps the von Neumann algebra ${\cal A}({\cal M})$ into its commutant. The second one means that $\Delta^{it}$ defines an automorphism of ${\cal A}({\cal M})$. This Theorem has far-reaching applications in Quantum Field Theory and Statistical Mechanics. \\\\As far as the Bell-CHSH inequality is concerned, we remind the important fact that the Tomita-Takesaki modular theory has been successfully employed for several spacetime regions as, for instance: wedge regions \cite{BW} and diamond regions \cite{Hislop:1981uh}. To be specific, in what follows we shall make always reference to the case of diamond regions, although similar results hold for wedge regions too. \\\\To proceed, we observe that the action of the operators $(J, \Delta)$ can be lifted into the space of test functions ${\cal m}(O)$ by means of \cite{Summers:1987fn,Summ,Summers:1987ze}
\begin{eqnarray} 
J\; e^{i \varphi(f)}\; J = e^{-i \varphi(jf)} \;, \nonumber \\
\Delta^{1/2} \; e^{i \varphi(f)}\; \Delta^{-1/2} = e^{i \varphi(\delta^{1/2} f)} \;. \label{lift}
\end{eqnarray}
Analogously to $(J, \Delta)$ , the operators $(j, \delta)$ are such thta 
\begin{eqnarray} 
s & =&  j \delta^{1/2} \;, \qquad s^2 =1 \;, \nonumber \\
j^2& =& 1 \;, \qquad j \delta^{1/2} j = \delta^{-1/2} \;, \nonumber \\
s^{\dagger} &= &j \delta^{-1/2} \;, \qquad \delta = s^\dagger s \;. \label{alsjd}
\end{eqnarray}
The operator $j$ is anti-unitary, while $\delta$ is positive and self-adjoint. In particular, as shown in  \cite{BW,Hislop:1981uh}, the spectrum of $\delta$ is the positive real line ${\mathbb{R}}_{+}$. Moreover, in the space of test functions, the operators $(s,s^{\dagger})$ act as projectors onto the space ${\cal M}$ and its symplectic complement ${\cal M}'$. In fact, it can be shown \cite{Guido:2008jk} that a test function $f$ belongs to ${\cal M}$ if and only if 
\begin{equation} 
s f = f \;. \label{sft}
\end{equation}
also, a test function $g$ belongs to ${\cal M}'$ if and only if 
\begin{equation}
s^{\dagger} g = g \;. \label{sdgt}
\end{equation}
This statement can be understood as follows. Let $(f,g)$ be two test functions obeying eqs.\eqref{sft},\eqref{sdgt}. Consider the inner product $\langle g|f\rangle$. From \eqref{sft},\eqref{sdgt} we have 
\begin{equation} 
\langle g |f \rangle = \langle s^{\dagger} g |f\rangle  = \langle s f| g\rangle = \langle f | g\rangle \;, \label{zz1}
\end{equation}
where use has been made of the anti-linearity of the operators $(s,s^\dagger)$. Therefore 
\begin{equation} 
\Im \langle g | f \rangle = \frac{1}{2} \Delta_{PJ}(g,f) = 0 \;, \label{zz2}
\end{equation}
so that $(g,f)$ are space-like. We are now ready to discuss the Bell-CHSH inequality. 
\subsection{The bell-CHSH inequality}\label{JB}
In order to study the Bell-CHSH inequality, we start from the correlation function 
\begin{equation} 
\langle 0|\; {\cal C}_i\;|0\rangle = \langle0|\; (A_i(f) + A_i(f'))B_i(g) + (A_i(f)-A_i(f') )B_i(g') \;|0\rangle \;, \label{CB}
\end{equation}
where $i=1,2,3$ labels the operators given in expressions \eqref{opftA},\eqref{opftB}. \\\\As outlined in \cite{Summers:1987fn,Summ,Summers:1987ze}, the test functions $(f,f')$ and $(g,g')$ needed for the evaluation of expression \eqref{CB} can be characterized by means of the modular operators $(s,j,\delta)$. From the knowledge of the spectrum of $\delta$ \cite{BW,Hislop:1981uh}, one  considers the spectral subspace $[\lambda^2-\varepsilon,\lambda^2+ \varepsilon] \in [0,1]$. Next, following \cite{Summers:1987fn,Summ,Summers:1987ze}, we pick up a normalized vector $\phi$ belonging to this subspace. Thus, for $(f,f')$ and $(g,g')$ we write 
\begin{eqnarray} 
f & =&  \eta (1+s) \phi \;, \qquad f'= \eta'(1+s) (i\phi) \;, \nonumber \\
g & = & \sigma j(1+s) \phi \;, \qquad g'= \sigma' j(1+s) (i\phi) \;, \label{TTtf}
\end{eqnarray} 
where $(\eta, \eta')$, $(\sigma,\sigma')$ are real arbitrary constants which, as much as the four Bell angles of Quantum Mechanics, can be chosen at our best convenience. \\\\Using 
\begin{equation} 
s^{\dagger} j (1+s) = j \delta^{-1/2} j (1+s) = \delta^{1/2} (1+s) = j (j\delta^{1/2}) (1+s) = j s(1+s) = j (1+s) \;, \label{pjsd}
\end{equation}
it follows that 
\begin{eqnarray} 
sf & =& f \;, \qquad s f'=f'\;, \nonumber \\
s^{\dagger} g &=& g \;, \qquad s^{\dagger} g'= g'\;. \label{ffpggp}
\end{eqnarray}
which imply thta $(f,f') \in {\cal M}$, while $(g,g')\in {\cal M}'$. As such, as needed, $(f,f')$ and $(g,g')$. are space-like. Moreover, recalling that $\phi$ belongs to the spectral subspace $[\lambda^2-\varepsilon, \lambda^2+\varepsilon]$, it turns out that \cite{Guimaraes:2024mmp}
\begin{eqnarray} 
||f||^2 & =& \langle f|f\rangle = \eta^2(1+\lambda^2) \;, \nonumber \\
||f'||^2 & = & \eta'^2 (1+\lambda^2) \;, \nonumber \\
||g||^2 & = & \sigma^2 (1+ \lambda^2) \;, \nonumber \\
||g'||^2 & = & \sigma'^2 (1+\lambda^2) \;, \nonumber \\
\langle f|g\rangle & =& 2 \eta \sigma \lambda \;, \nonumber \\
\langle f'| g'\rangle & = & 2 \eta'\sigma'\lambda \;, \nonumber \\
\langle f|g' \rangle & = & \langle f'|g\rangle = 0 \;. \label{innffpggp}
\end{eqnarray}
These relations are consequence of the fact that $\phi$ and $j\phi$ belong to different spectral subspaces. In fact 
\begin{equation} 
\delta (j\phi) = j j \delta j \phi = \j \delta ^{-1} \phi \;, \label{subsp}
\end{equation}
which shows that the modular conjugation $j$ exchanges the spectral subspace $[\lambda^2 -\varepsilon, \lambda^2 + \varepsilon]$ into 
$[1/\lambda^2 -\varepsilon, 1/\lambda^2 + \varepsilon]$. As a consequence, $\phi$ and $j\phi$ are orthogonal, {\it i.e.} 
\begin{equation} 
\langle \phi | j\phi \rangle = 0 \;, \label{phijphi} 
\end{equation}
from which eqs.\eqref{innffpggp} are derived. \\\\Therefore, for the Bell-CHSH inequality one gets 
\begin{eqnarray} 
\langle 0|\; {\cal C}_i\;|0\rangle & = & \int_{-\infty}^{\infty} dk dp \;{\hat \sigma}_i(k) {\hat \sigma}_i(p) \;e^{-\frac{1}{2}(\eta^2(1+\lambda^2) +\sigma^2(1+\lambda^2) + 4pk \eta \sigma \lambda)} \nonumber \\
& + &  \int_{-\infty}^{\infty} dk dp \;{\hat \sigma}_i(k) {\hat \sigma}_i(p) \;e^{-\frac{1}{2}(\eta'^2(1+\lambda^2) +\sigma^2(1+\lambda^2) )} \nonumber \\
& + &  \int_{-\infty}^{\infty} dk dp \;{\hat \sigma}_i(k) {\hat \sigma}_i(p) \;e^{-\frac{1}{2}(\eta^2(1+\lambda^2) +\sigma'^2(1+\lambda^2) )} \nonumber \\
&-& \int_{-\infty}^{\infty} dk dp \;{\hat \sigma}_i(k) {\hat \sigma}_i(p) \;e^{-\frac{1}{2}(\eta'^2(1+\lambda^2) +\sigma'^2(1+\lambda^2) + 4pk \eta' \sigma \lambda)} \;, \label{BTTT}
\end{eqnarray}
where the functions ${\hat \sigma}_i(k), i=1,2,3$ are given in eq.\eqref{ftkex}. This concise and elegant expression shows the usefulness of the Tomita-Takesaki modular theory. It gives rise to quite large violations of the Bell-CHSH inequality, close to Tsirelson's bound $2 \sqrt{2}$. For instance, in the case of the operators $(A_2,B_2)$ one gets 
\begin{equation} 
\langle 0|\; {\cal C}_2 \;|0\rangle = 2.723 \;, \label{violC2}
\end{equation}
for 
\begin{eqnarray} 
\eta & = & 0.024 \;, \qquad \eta'=4.732 \;, \qquad \sigma = 0.086 \;,\nonumber \\
\sigma' & =  & 9.307 \;, \qquad \lambda = 0.884 \;. \label{valuesp}
\end{eqnarray} 
Similar results are obtained for the other operators $(A_1, B_1)$, $(A_3, B_3)$.

\section{The numerical setup.}\label{nm}

\begin{figure}[t!]
	\begin{minipage}[b]{0.6\linewidth}
		\includegraphics[width=\textwidth]{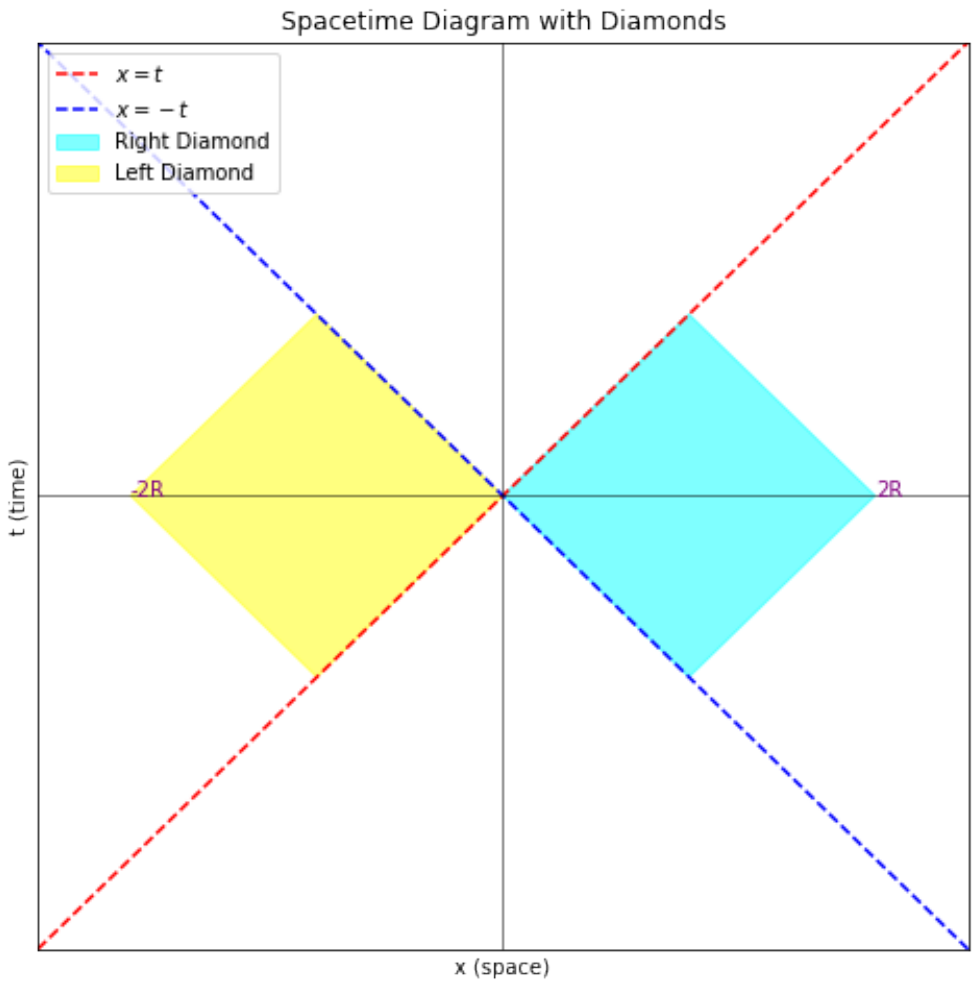}
	\end{minipage} \hfill
\caption{Causal tangent double diamond regions.  The parameter $R$ determines the size of the diamonds.}
	\label{diamonds}
	\end{figure}

\begin{figure}[t!]
	\begin{minipage}[b]{0.6\linewidth}
		\includegraphics[width=\textwidth]{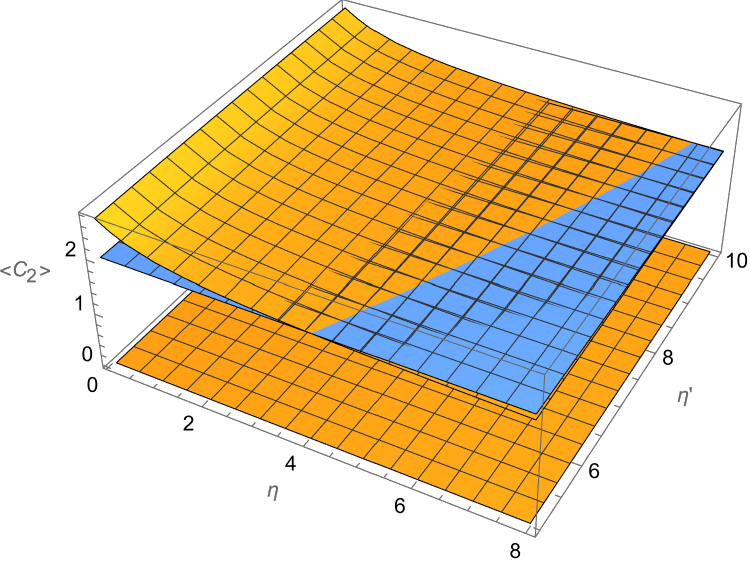}
	\end{minipage} \hfill
\caption{The Bell-CHSH correlation function for the operators $(A_2,B_2)$, eqs.\eqref{opftA},\eqref{opftB}, as a function of the parameters $(\eta, eta')$, for $(\sigma=0.011, \sigma'=2.102, \lambda= 0.884)$. The blue surface denotes the classical bound: 2. The orange surface above the blue one shows the region in which the violation occurs.}
	\label{c2}
	\end{figure}

\begin{figure}[t!]
	\begin{minipage}[b]{0.6\linewidth}
		\includegraphics[width=\textwidth]{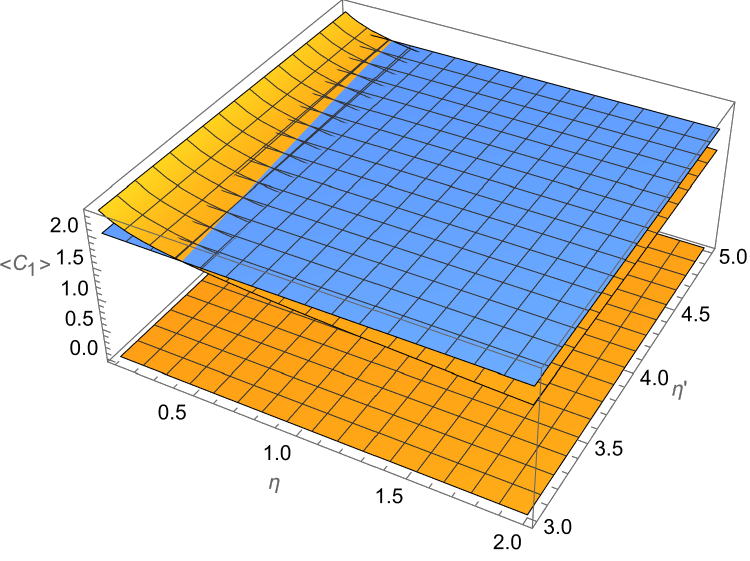}
	\end{minipage} \hfill
\caption{The Bell-CHSH correlation function for the operators $(A_1,B_1)$, eqs.\eqref{opftA},\eqref{opftB}, as a function of the parameters $(\eta, eta')$, for $(\sigma=0.144, \sigma'=8.714, \lambda= 0.884)$. The blue surface denotes the classical bound: 2. The orange surface above the blue one shows the region in which the violation occurs.}
	\label{c1}
	\end{figure}

\begin{figure}[t!]
	\begin{minipage}[b]{0.6\linewidth}
		\includegraphics[width=\textwidth]{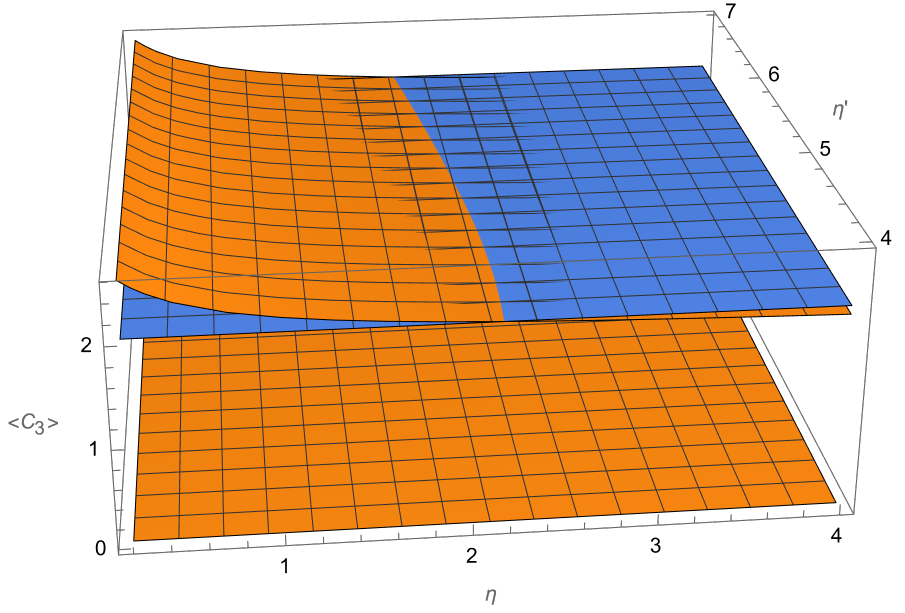}
	\end{minipage} \hfill
\caption{The Bell-CHSH correlation function for the operators $(A_3,B_3)$, eqs.\eqref{opftA},\eqref{opftB}, as a function of the parameters $(\eta, eta')$, for $(\sigma=0.104, \sigma'=,8.784, \lambda= 0.884)$. The blue surface denotes the classical bound: 2. The orange surface above the blue one shows the region in which the violation occurs.}
	\label{c3}
	\end{figure}

Having discussed the Bell-CHSH inequality within the framework of the Tomita-Takesaki modular theory, we proceed by presenting the construction of a numerical setup, the aim being that of providing a comparison with the previous results, besides giving a second way of facing the Bell-CHSH inequality. \\\\To start, we need to be more precise about the shape and size of the spacetime regions which will be considered. For that, we shall take two causal tangent diamonds, as illustrated in Fig.\eqref{diamonds}
\\\\From a numerical side, the explicit construction of a set of test functions $(f,f')$ and $(g,g')$ fulfilling the conditions \eqref{innffpggp} derived from the Tomita-Takesaki modular theory is very challenging. As one can figure out, this is is due to the difficulties of constructing a numerical version of the modular operators $(j,\delta)$. To overcome this point, we proceed to achieve an optimal set of test functions which will allow to stay as close as possible to those of the Tomita-Takesaki theory. As such, these test functions will guarantee that the violations of the Bell-CHSH inequality will be in good agreement with those of the modular theory. \\\\Let us focus first on the test functions $(f,f')$, supported in the right diamond specified by 
\begin{equation} 
|x-R| + |t| \le R \;, \label{rd}
\end{equation} 
where the parameter $R$ determines the size of the diamond. \\\\For $(f,f')$ we set 
 \begin{equation} 
 f(t,x) = \eta (1+\lambda^2)^{1/2} \frac{{\hat f}(t,x)}{\sqrt{H({\hat f},{\hat f})}}\;, \qquad  f'(t,x) = \eta' (1+\lambda^2)^{1/2} \frac{{\hat f'}(t,x)}{\sqrt{H({\hat f'},{\hat f'})}}\;, \label{nmffp}
 \end{equation}
 where $({\hat f}(t,x),{\hat f'}(t,x))$ are supported in the right diamond, namely 
\begin{align}
{\hat f}(t,x) =\left\{
    \begin {aligned}
         & e^{-\frac{a}{R^2 - (|x-R| +|t|)^2}} \quad & |x-R| + |t| \le R   \\
         & 0 \quad & {\rm elsewhere}                   
    \end{aligned}
\right. \label{ff}
\end{align} 
\begin{align}
{\hat f}'(t,x) =
      \left\{
    \begin {aligned}
         & e^{-\frac{a}{R^2 - (|x-R| +|t|)^2}} \quad & |x-R| + |t| \le R   \\
         & 0 \quad & {\rm elsewhere}                   
    \end{aligned}
\right. \label{ffp}
\end{align}
with $(a,a')$ are constant parameters and 
\begin{eqnarray} 
H({\hat f},{\hat f}) & = & \langle {\hat f}| {\hat f}\rangle =  \int d^2x \; d^2x' \; {\hat f}(x) H(x-x') {\hat f}(x') \;, \nonumber \\
H({\hat f'},{\hat f'}) & = & \langle {\hat f'}| {\hat f'}\rangle =  \int d^2x \; d^2x' \; {\hat f'}(x) H(x-x') {\hat f'}(x') \;. \label{HHH}
\end{eqnarray}
Similarly, for $(g,g')$ one sets
\begin{equation} 
g(t,x) = \sigma (1+\lambda^2)^{1/2} \frac{{\hat g}(t,x)}{\sqrt{H({\hat g},{\hat g})}}\;, \qquad  g'(t,x) = \sigma' (1+\lambda^2)^{1/2} \frac{{\hat g'}(t,x)}{\sqrt{H({\hat g'},{\hat g'})}}\;, \label{nmggp}
\end{equation}
where $({\hat g}, {\hat g'})$ are supported in the left diamond:
\begin{align}
{\hat g}(t,x) = 
\left\{
    \begin {aligned}
         & e^{-\frac{b}{R^2 - (|x+R| +|t|)^2}} \quad & |x+R| + |t| \le R   \\
         & 0 \quad & {\rm elsewhere}                   
    \end{aligned}
\right. \label{gg}
\end{align}
and 
\begin{align}
{\hat g'}(t,x) = 
      \left\{
    \begin {aligned}
         & e^{-\frac{b'}{R'^2 - (|x+R'| +|t|)^2}} \quad & |x+R'| + |t| \le R'   \\
         & 0 \quad & {\rm elsewhere}                   
    \end{aligned}
\right. \label{ggp}
\end{align}
At this stage, the parameters $(a,a')$ and $(b,b')$ are free parameters. Aa already underlined, these parameters will be determined, together with $R$,  in such a way to stay as close as possible to the results established in the previous sections. \\\\For the inner products we get now 
\begin{eqnarray} 
|| f||^2 & = & \eta^2 (1+ \lambda^2) \;, \qquad ||f'||^2 = \eta'^2 (1+\lambda^2) \;, \nonumber \\
||g||^2 & = & \sigma^2 (1+ \lambda^2) \;, \qquad ||g'||^2 = \sigma'^2 (1+\lambda^2) \;, \nonumber \\
\langle f | g \rangle & = & \eta \sigma (1+ \lambda^2) \frac{ H({\hat f},{\hat g}) }{ \sqrt{H({\hat f},{\hat f}) H({\hat g},{\hat g})}} \;, \nonumber \\
\langle f | g' \rangle & = & \eta \sigma' (1+ \lambda^2) \frac{ H({\hat f},{\hat g'}) }{ \sqrt{H({\hat f},{\hat f}) H({\hat g'},{\hat g})}} \;, \nonumber \\
\langle f' | g \rangle & = & \eta' \sigma (1+ \lambda^2) \frac{ H({\hat f'},{\hat g}) }{ \sqrt{H({\hat f'},{\hat f'}) H({\hat g},{\hat g})}} \;, \nonumber \\
\langle f' | g' \rangle & = & \eta' \sigma' (1+ \lambda^2) \frac{ H({\hat f'},{\hat g'}) }{ \sqrt{H({\hat f'},{\hat f'}) H({\hat g'},{\hat g})}} \;, \nonumber \\
\end{eqnarray}
Therefore, for the Bell-CHSH inequality, one has 
\begin{eqnarray} 
\langle 0|\; {\cal C}_i\;|0\rangle & = & \int_{-\infty}^{\infty} dk dp \;{\hat \sigma}_i(k) {\hat \sigma}_i(p) \;e^{-\frac{1}{2}(\eta^2(1+\lambda^2) +\sigma^2(1+\lambda^2) + 2pk \eta \sigma (1+\lambda^2) \alpha )} \nonumber \\
& + &  \int_{-\infty}^{\infty} dk dp \;{\hat \sigma}_i(k) {\hat \sigma}_i(p) \;e^{-\frac{1}{2}(\eta'^2(1+\lambda^2) +\sigma^2(1+\lambda^2)+ 2kp \eta'\sigma (1+\lambda^2) \beta )} \nonumber \\
& + &  \int_{-\infty}^{\infty} dk dp \;{\hat \sigma}_i(k) {\hat \sigma}_i(p) \;e^{-\frac{1}{2}(\eta^2(1+\lambda^2) +\sigma'^2(1+\lambda^2) + 2 kp \eta \sigma'(1+\lambda^2) \gamma )} \nonumber \\
&-& \int_{-\infty}^{\infty} dk dp \;{\hat \sigma}_i(k) {\hat \sigma}_i(p) \;e^{-\frac{1}{2}(\eta'^2(1+\lambda^2) +\sigma'^2(1+\lambda^2) + 2pk \eta' \sigma (1+\lambda^2) \delta )} \;. \label{BNNN}
\end{eqnarray}
By comparing expression \eqref{BNNN} with that obtained from the Tomita-Takesaki theory, eq.\eqref{BTTT}, one sees that the parameters $(\alpha, \beta,  \gamma, \delta)$ 
\begin{eqnarray} 
\alpha & = &  \frac{ H({\hat f},{\hat g}) }{ \sqrt{H({\hat f},{\hat f}) H({\hat g},{\hat g})}} \; \qquad \beta =  \frac{ H({\hat f'},{\hat g}) }{ \sqrt{H({\hat f'},{\hat f}) H({\hat g},{\hat g})}} \;, \nonumber \\
\gamma & = &  \frac{ H({\hat f},{\hat g'}) }{ \sqrt{H({\hat f},{\hat f}) H({\hat g'},{\hat g'})}} \;, \qquad \delta =  \frac{ H({\hat f'},{\hat g'}) }{ \sqrt{H({\hat f'},{\hat f'}) H({\hat g'},{\hat g'})}} \;, \label{cabgd}
\end{eqnarray} 
give a measure of how different the two expressions are. In particular, in the case of the modular theory, we have 
\begin{equation} 
\alpha_{TT} = \frac{2 \lambda}{1+\lambda^2} \;, \qquad \beta_{TT}=0 \;, \qquad \gamma_{TT}=0 \;, \qquad \delta_{TT} =  \frac{2 \lambda}{1+\lambda^2} \;, \label{cTTT}
\end{equation}
which, using the value $\lambda=0.884$ employed in eq.\eqref{valuesp}, become 
\begin{equation} 
\alpha_{TT} = 0.992 \;, \qquad \beta_{TT}=0 \;, \qquad \gamma_{TT}=0 \;, \qquad \delta_{TT} = 0.992 \;. \label{cTTTc}
\end{equation}
We attempted at reproducing these numerical values by searching for  suitable values of the parameters $(a,a',b,b',R)$ entering the test functions $(f,f',g,g')$. This has required a detailed study of the highly complex four-dimensional spacetime integrals \eqref{HHH}. Such a study has been performed using Mathematica. The above mentioned integrals have been faced by employing the MonteCarlo and Quasi-MonteCarlo methods. After a rather lengthy analysis, we have been able to fix the parameters $(a,a',b,b', R)$ in such a way to remain sufficiently close to expressions \eqref{cTTTc}, namely: 
\begin{equation} 
\alpha = 0.944 \;, \qquad \beta=0 \;, \qquad \gamma=0.732 \;, \qquad \delta = 0.906 \;. \label{cTTTn}
\end{equation}
Even though the values \eqref{cTTTn} do not coincide exactly with those of eq.\eqref{cTTTc}, they give rise to rather large violations of the Bell-CHSH inequality, which compare well with those obtained from the Tomita-Takesaki modular theory. This feature is nicely illustrated in Fig.\eqref{c2}, where the violation of the Bell-CHSH as function pf the parameters $(\eta, \eta')$ is reported. The blue surface denotes the classical bound: 2, while the orange surface above the blue one shows the region in parameter space in which the violation occurs. In particular we notice that 
\begin{eqnarray} 
\langle {\cal C}_2\rangle_{\rm num} & =&  2.752 \;,  \qquad 
\eta  =  0.024 \;, \qquad \eta'=4.732 \;, \qquad \; \nonumber \\
\sigma &= & 0.086 \; \qquad 
\sigma'  =   9.307 \;, \qquad \lambda = 0.884 \;, \label{valuenum}
\end{eqnarray} 
which is in agreement with the value reprted from the modular theory, eq.\eqref{valuesp}.\\\\Let us also show, for completeness, the plots of the Bell-CHSH correlation functions corresponding to the operators $(A_1,B_1)$ and $(A_3, B_3)$, displayed in Figures \eqref{c1},\eqref{c3}. 
. 
One sees that, in all three cases, there exist rather large regions in parameter space for which sensible violations of the Bell-CHSH inequality occurr. \\\\Let us end with a remark on the mass parameter $m$, which appears in the Hadamard distribution, eq.\eqref{PJH}. As far as the diamond regions are considered, it should be reminded that the modular theory has been established for massless theories \cite{Hislop:1981uh}. This feature is deeply connected with the fact that the symmetry transformations which leave the diamonds invariant are the conformal transformations. Accordingly, in the numerical setup we have adopted a very small mass, $m \sim 10^{-8}$,  playing the role of an infrared cutoff.

\section{Conclusion}\label{Ccc}

In this work we have pursued the study of the violation of the Bell-CHSH inequality in the vacuum state of a relativistic scalar Quantum Field Theory. \\\\The main results which have been achieved can be summarized as: 
\begin{itemize} 
\item construction of a useful set of bounded Hermitian operators for the Bell-CHSH correlation functions. The reasoning here has been that of making use of real functions having known Fourier transformation. This feature has enabled us to introduce a set of operators, eqs.\eqref{opftA},\eqref{opftB}, whose correlation functions can be evaluated by means of the algebra of the unitary Weyl operators. 
\item a helpful parametrization, eq.\eqref{BNNN}, of the Bell-CHSH correlator has been introduced. Such a parametrization has enabled an efficient comparison between the theoretical framework based on the Tomita-Takesaki modular theory and the explicit numerical setup presented in Sect.\eqref{nm}, thus establishing a bridge between the Theorems proven in \cite{Summers:1987fn,Summ,Summers:1987ze} and an explicit computational framework. 
\item three examples of bounded Hermitian operators have been investigated, eqs.\eqref{opftA},\eqref{opftB}. All operators display rather large violations of the Bell-CHSH inequality. 
\end{itemize}

\section*{Acknowledgments}
The authors would like to thank the Brazilian agencies CNPq, CAPES end FAPERJ for financial support.  S.P.~Sorella, I.~Roditi, and M.S.~Guimaraes are CNPq researchers under contracts 301030/2019-7, 311876/2021-8, and 309793/2023-8, respectively.

\appendix

\section{The massive real scalar field in 1+1 Minkowski spacetime}\label{appA}

The massive real scalar field in 1+1-dimensional Minkowski spacetime has the plane-wave expansion:
\begin{equation} \label{qf}
\varphi(t,x) = \int \! \frac{d k}{2 \pi} \frac{1}{2 \omega_k} \left( e^{-ik_\mu x^\mu} a_k + e^{ik_\mu x^\mu} a^{\dagger}_k \right), 
\end{equation} 
where $\omega_k  = k^0 = \sqrt{k^2 + m^2}$. For the canonical commutation relations, one has 
\begin{align}
[a_k, a^{\dagger}_q] &= 2\pi \, 2\omega_k \, \delta(k - q), \\ \nonumber 
[a_k, a_q] &= [a^{\dagger}_k, a^{\dagger}_q] = 0. 
\end{align}
It is a well-known fact that quantum fields must be considered as operator-valued distributions \cite{Haag:1992hx}. Consequently, they need to be smeared to produce well-defined operators that act on the Hilbert space, {\it i.e.}
\begin{align} 
\varphi(h) = \int \! d^2x \; \varphi(x) h(x) \;, \label{smmd}
\end{align}
where $h$ is a real smooth test function with compact support. With the smeared fields, the Lorentz-invariant inner product is introduced by means of the two-point  smeared Wightman function
\begin{align} \label{InnerProduct}
\langle f \vert g \rangle &= \langle 0 \vert \varphi(f) \varphi(g) \vert 0 \rangle =  \frac{i}{2} \Delta_{PJ}(f,g) +  H(f,g) \;, 
\end{align}
where $f$ and $g$ are also real smooth test functions with compact support, and $ \Delta_{PJ}(f,g)$ and $H(f,g)$ are the smeared versions of the Pauli-Jordan and Hadamard expressions
\begin{align}
\Delta_{PJ}(f,g) &=  \int \! d^2x d^2y f(x) \Delta_{PJ}(x-y) g(y) \;,  \nonumber \\
H(f,g) &=  \int \! d^2x d^2y f(x) H(x-y) g(y)\;. \label{mint}
\end{align}
Here, $\Delta_{PJ}(x-y)$ and $H(x-y)$ are given by
\begin{eqnarray} 
\Delta_{PJ}(t,x) & =&  -\frac{1}{2}\;{\rm sign}(t) \; \theta \left( \lambda(t,x) \right) \;J_0 \left(m\sqrt{\lambda(t,x)}\right) \;, \nonumber \\
H(t,x) & = & -\frac{1}{2}\; \theta \left(\lambda(t,x) \right )\; Y_0 \left(m\sqrt{\lambda(t,x)}\right)+ \frac{1}{\pi}\;  \theta \left(-\lambda(t,x) \right)\; K_0\left(m\sqrt{-\lambda(t,x)}\right) \;, \label{PJH}
\end{eqnarray}
where 
\begin{equation} 
\lambda(t,x) = t^2-x^2 \;, \label{ltx}
\end{equation}
and $(J_0,Y_0,K_0)$ are Bessel functions, while $m$ is the mass parameter. \\\\Both the Hadamard and Pauli-Jordan distributions are Lorentz-invariant. Notably, the Pauli-Jordan distribution, \(\Delta_{PJ}(x)\), encodes relativistic causality, as it vanishes outside the light cone. Furthermore, \(\Delta_{PJ}(x)\) and the Hadamard distribution, \(H(x)\), exhibit distinct symmetry properties: \(\Delta_{PJ}(x)\) is odd under the transformation \(x \to -x\), whereas \(H(x)\) is even. When expressed in terms of smeared fields, the commutator of the field operators takes the form 
\[
\left[\varphi(f), \varphi(g)\right] = i \Delta_{PJ}(f, g).
\]
Within this framework, causality is elegantly encapsulated by the condition $\left[\phi(f), \phi(g)\right] = 0,$
whenever the supports of $f$ and $g$ are spacelike separated.


\end{document}